# What happens to the chemical equilibrium in case of a reference frame change

*Friedrich Herrmann and Michael Pohlig, Abteilung für Didaktik der Physik, Karlsruhe Institute of Technology, D-76128 Karlsruhe*

*E-mail: f.herrmann@kit.edu, michael.pohlig@kit.edu*

*Abstract:* It is shown that upon a reference frame change, the chemical potential transforms in the same way as temperature and electric potential. Consequently, during a reference frame change, a chemical equilibrium remains a chemical equilibrium.

## 1. Introduction

When addressing reference frame changes, the primary concern is the behavior of physical quantities from mechanics or electrodynamics, which is particularly significant for particle physics and cosmology. However, not only mechanical and electrical quantities transform during a reference frame change, but also thermodynamic and chemical quantities. In the following discussion, we will explore these transformations.

Thereby, we might answer a question that readers do not initially have. Why do we do it anyway? Because it can be intriguing to encounter new questions and, naturally, to find their answers. We formulate our question using a brief story that comes across as a paradox.

Bob is on the terrace of the station restaurant, and Alice is on a passing train. It's a hot day. Bob has a glass of iced water in front of him, with the ice finely crushed, resulting in a mixture at 0°C. Alice has a similar drink. Bob is puzzled that the ice in Alice's glass is floating in liquid water, because he had learned that according to the theory of relativity the temperature in a moving reference frame is lower than in a stationary one [1, 2, 3, 4]. Therefore, he believes Alice's drink should be below 0°C and frozen. But Alice is equally surprised, expecting Bob's drink to be frozen. Both assume that the melting temperature is an intrinsic property that cannot change with a reference frame change. End of the story.

We aim to provide an explanation that goes back to the cause of the coexistence of two phases. We discuss the phase transition between solid and liquid as an example, though the results apply to any phase transition or chemical reaction, i.e., any chemical equilibrium.

## 2. Coexistence of two phases as chemical equilibrium

One can describe the melting point such as that of water via an observation. When the temperature of ice (solid water) is increased from a low value and the melting temperature or "melting point" is reached, the ice begins to melt. Similarly, starting from a high temperature, the water begins to solidify when the melting temperature is reached. At the melting point, the solid and liquid phases coexist.

In this description, nothing is said about the reason for this coexistence. We now will give such a reason.

The solid and liquid phases coexist when the chemical potentials of both phases are equal. Below the melting point, the chemical potential of the liquid phase is higher than that of the solid phase, and above it the reverse is true (Fig. 1) [5, 6]. The difference in the chemical potentials can be interpreted as a driving force for the phase transition.

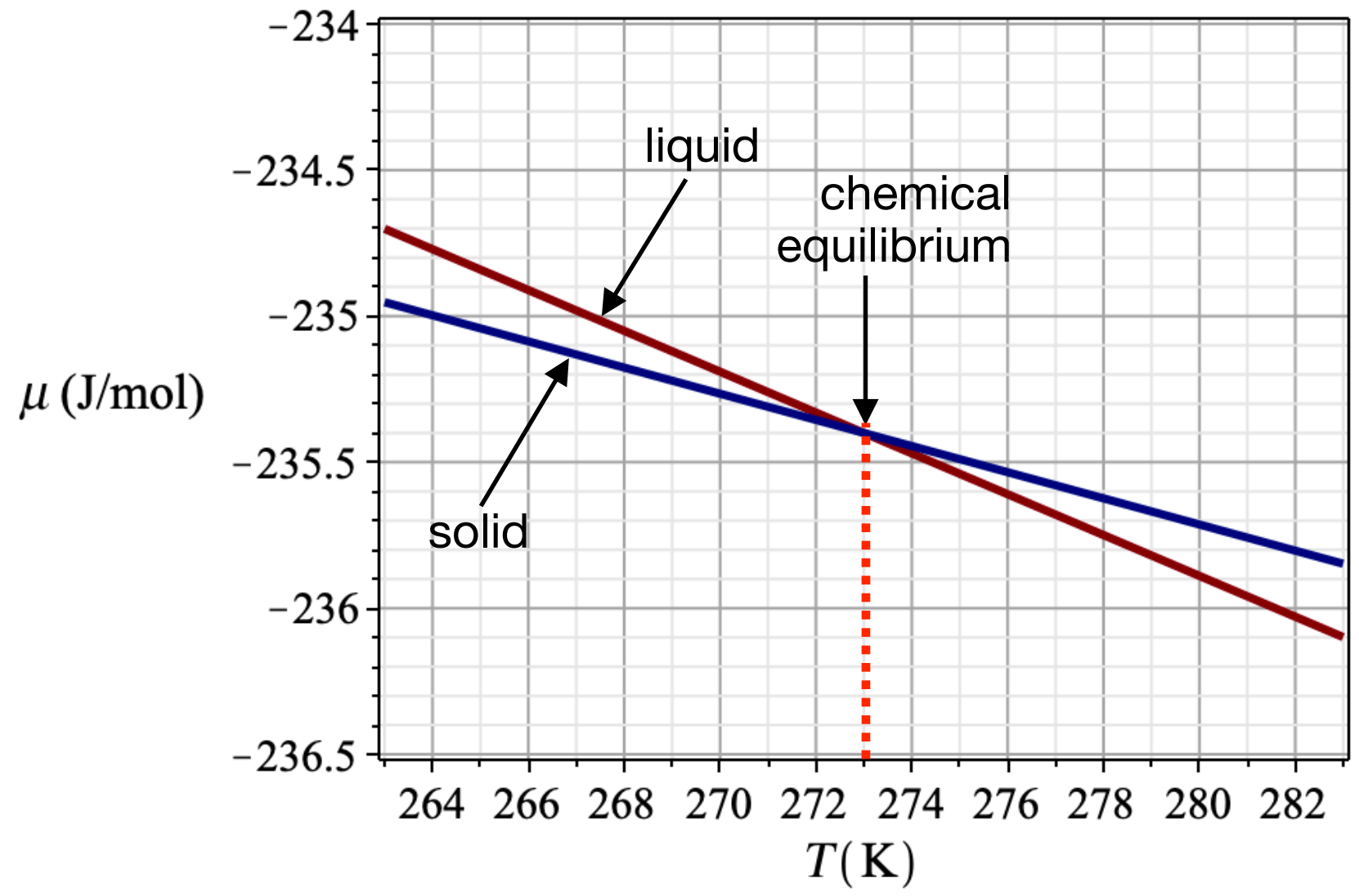

Fig. 1: Chemical potential of the solid and liquid phases of water at a pressure of 1 bar as a function of temperature. When both values are equal, the two phases coexist.

To determine what happens to the melting temperature when the reference frame is changed, we have to investigate how the chemical potential transforms upon a reference frame change.

### 3. The dependence of the chemical potential on the reference frame

We begin with Gibbs' fundamental equation, written in a general form [7, 8]:

$$dE = TdS - pdV + \mu dn + UdQ + \vec{v} \cdot d\vec{p} + \vec{\omega} \cdot d\vec{L} + \ldots \tag{1}$$

The equation tells us how the energy of a system changes when another extensive quantity of the system (entropy $S$, volume $V$, amount of substance $n$, electric charge $Q$, momentum $\vec{p}$, angular momentum $\vec{L}$, etc.) is changed. The pre-factors (absolute temperature $T$, pressure $p$, chemical potential $\mu$, electric potential or potential difference $U$, velocity $\vec{v}$, angular velocity $\vec{\omega}$, etc.) are intensive quantities, “energy-conjugated” to the respective extensive quantities.

We now consider a process where the quantity $n$ (amount of substance) is supposed to change in the system at rest (denoted by subscript 0). Equation (1) reduces to

$$dE_0 = \mu_0 dn_0 \tag{2}$$

In order to change the quantity “amount of substance”, we have to add matter to our system. However, this is inevitably linked to a convective entropy supply. We eliminate this contribution to the energy change by removing the corresponding amount of entropy from the system by heat conduction, i.e. not convectively.

We now consider the same process in a moving reference frame. The transferred energy corresponding to the term $\mu dn$ is equivalent to transferred mass, and this means that in the moving reference frame momentum is also transferred. The corresponding Gibbs fundamental equation must now include the term $\vec{v} \cdot d\vec{p}$:

$$dE = \mu dn + \vec{v} \cdot d\vec{p}, \tag{3}$$

Here $d\vec{p}$ is the momentum supplied to the system and $\vec{v}$ the velocity of the moving reference frame relative to the reference frame at rest.

We write

$$d\vec{p} = d(m\vec{v}) = md\vec{v} + \vec{v}dm$$

Since

$$d\vec{v} = 0,$$

we get

$$d\vec{p} = \vec{v}dm$$

Using

$$dm = \frac{dE}{c^2}$$

we obtain

$$d\vec{p} = \frac{\vec{v}dE}{c^2}$$

and finally

$$\vec{v}d\vec{p} = \frac{v^2}{c^2}dE$$

Substitution into equation (3) results in

$$dE = \mu dn + \frac{v^2}{c^2}dE$$

and thus

$$dE(1 - \frac{v^2}{c^2}) = \mu dn \tag{4}$$

In the following, we abbreviate as usual:

$$\beta = \frac{v}{c}$$

The amount of substance is Lorentz-invariant, so it is

$$dn = dn_0$$

and the energy transforms according to

$$dE = \frac{dE_0}{\sqrt{1-\beta^2}}$$

Thus, equation (4) becomes

$$dE_0\sqrt{1-\beta^2} = \mu dn_0$$

With equation (2) we finally obtain

$$\mu = \mu_0\sqrt{1-\beta^2} \tag{5}$$

The chemical potential in the moving reference frame, seen from the rest frame, is reduced by the factor

$$\sqrt{1-\beta^2}\ ,$$

the same factor that is known from other transformations, such as that of the length (length contraction).

The simple relationship of equation (5) arises because the amount of substance is Lorentz-invariant. However, other extensive quantities, like electric charge and entropy, are also Lorentz-invariant. Consequently, the corresponding intensive quantities transform similarly to the chemical potential. We thus get, as a kind of by-catch:

$$T = T_0\sqrt{1-\beta^2} \qquad (6)$$

and

$$U = U_0\sqrt{1-\beta^2} \qquad (7)$$

The first relationship (6) was already derived shortly after the publication of the special theory of relativity [1,2,3], but was later questioned. In some papers, another relationship was derived [9], namely

$$T = \frac{T_0}{\sqrt{1-\beta^2}}$$

This result is obtained if the term $\vec{v}\cdot d\vec{p}$ in equation (3) is not taken into account.

There are other authors [10] who believe that the transformation law of the temperature is a subject of choice. Our thermodynamic derivation shows that such a choice does not exist - just as it does not exist for the generally accepted transformation equation for the electric potential.

Equation (7) can also be derived in a different way. Consider a capacitor that moves orthogonally to its plates relative to the rest frame. The electric field component in this direction is invariant, the distance between the plates decreases by $\sqrt{1-\beta^2}$. Thus the voltage decreases by the same factor.

## 4. Equilibria

Back to our phase transitions. The chemical potential transforms according to equation (5), regardless of the substance involved and regardless of the phase in which the substance exists. Thus, if the chemical potentials of two substances or two phases are equal in the rest frame, they remain equal in the moving frame, since both are multiplied by the same factor.

The same applies to thermal equilibria because of equation (6) and to electrical equilibria because of equation (7), and it also applies if the sum of several intensive variables determines an equilibrium.

The best-known example of such a combined equilibrium is the electrochemical equilibrium. For example consider a currentless p-n junction in a semiconductor. The electric current density is zero at every point, although there is an electric potential gradient, i.e. a drive for a current of electric charge. However, this is balanced by a gradient of the che-mical potential. The gradient of the sum of the two potentials, i.e. the gradient of the electrochemical potential and thus the net drive, is zero [11]. We have local “electrochemical equilibrium”. We now describe the system in motion. The chemical potential and thus its gradient now has a different value. Does this result in a deviation from equilibrium? No, because the electric potential (and therefore also its gradient) changes by the same factor as the chemical potential.

## 4. Conclusion

The chemical potential transforms in the same way as temperature and electric potential upon a reference frame change. Therefore, a state of equilibrium (thermal, chemical, electrical, electrochemical and others) remains a state of equilibrium during a reference frame change.